\documentclass{emulateapj}

\usepackage{graphicx}
\usepackage{graphicx,epsfig}
\usepackage{bm}
\usepackage{latexsym,amssymb,amsmath,float}

\newcommand{\hmpc}{h^{-1}\mathrm{Mpc}}

\newcommand{\kms}{{\,{\rm km}\,{\rm s}^{-1}}}
\newcommand{\Omegam}{\Omega_{m}}

\bibpunct[,]{(}{)}{;}{a}{}{,}
\begin{document}

\title{From Finance to Cosmology: The Copula of Large-Scale Structure}

\author{Robert~J.~Scherrer, Andreas~A.~Berlind, Qingqing~Mao, Cameron~K.~Mcbride}

\affil{Department of Physics and Astronomy, Vanderbilt University,
Nashville, TN  37235}

\begin{abstract}
Any multivariate distribution can be uniquely decomposed
into marginal (1-point) distributions, and a function called the copula, which
contains all of the information on correlations between the distributions.
The copula
provides an important new methodology for analyzing the density field in large-scale
structure.
We derive the empirical 2-point copula for the evolved dark matter density field.
We find that this empirical copula is well-approximated by a Gaussian copula.  We consider
the possibility that
the full $n$-point copula is also Gaussian and describe some of the consequences of this
hypothesis.  Future directions for investigation are discussed.
\end{abstract}



\section{Introduction} \label{sec:intro}

The standard model for the formation of large-scale structure assumes that the
universe at high redshift contained a dark matter density field characterized
by a
multivariate Gaussian distribution.  This density field evolved,
under the action of gravity, into a highly non-Gaussian dark matter density
field, with the present-day observed distribution of galaxies tracing (in a
biased fashion) the underlying dark matter.

Many tools have been developed to characterize the final evolved distribution of
matter.  The most widely used are the $n$-point correlation functions
(Peebles 1980).  When
applied to a discrete density field (such as the observed galaxy distribution)
they give the probability (in excess of random) of observing galaxies at a set of $n$ points in a
fixed geometrical configuration relative to each other.  For a continuous
density field (such as the theoretical dark matter distribution) these $n$-point
functions can be expressed in terms of the density field measured at $n$
points at a fixed relative separation.  A knowledge of all of the
correlation functions up to arbitrarily large $n$ completely characterizes
a given density field or galaxy distribution.

The problem is that in practice, it is impossible to measure
correlation functions to arbitrarily high order.  The two-point correlation function is
known to very high accuracy, and the three-point function of the distribution
of galaxies is also well measured.  However, precise measurements of the 
four-point correlation function or any higher orders are difficult to impossible for 
current data. 
Although the two- and three-point correlation functions provide a great 
deal of information about the galaxy distribution, we are left with an incomplete 
characterization of this distribution.

Attempts have been made, therefore, to slice the information contained in the
density field (or in the distribution of galaxies) in different ways.
For example, the void probability function (White 1979, Fry, et al. 1988) mixes
together information from correlation functions of all orders, as do
percolation statistics (Zel'dovich 1982; Shandarin 1983; Sahni et al. 1997).
Similarly,
the 1-point probability distribution function (PDF) has been widely explored
(Coles \& Jones 1991; Kofman et al. 1994; Protogeros \& Scherrer 1997;
Scherrer \& Gaztanaga 2001; Lam \& Sheth 2008); it also samples the information in 
the density field in a different way from the correlation functions.  However, none of 
these statistics provides a complete description of the density field; 
they all sample only part of the information.

In the case of the 1-point PDF, however, it is possible to introduce a new
statistical tool, the copula, which provides the rest of the information contained in
the density field.  The copula and the 1-point PDF together completely characterize
the density distribution, and this decomposition is unique for any multivariate density field.
Roughly speaking, the copula indicates how the 1-point PDFs are joined together
to give the $n$-point PDF.

The copula was first defined and characterized by Sklar (1959) and it has been
most widely applied in the field of mathematical finance.  In fact, misuse of the
copula has been blamed for the recent meltdown in the mortgage-backed securities industry.
The copula has been used in various areas of engineering, especially hydrology (Genest
and Favre 2007), but
it has not been widely applied in astronomy or astrophysics (although see the recent
papers by Jiang, et al. 2009 and Benabed, et al. 2009).  To our knowledge, this paper represents the first
application to the analysis of large-scale structure.

In the next section, we review the definition and properties of the copula.
In \S 3, we apply the copula methodology to a simulated dark matter density
field in the standard $\Lambda$CDM cosmology.  We find that the 2-point copula
of the evolved density field is well-approximated by a Gaussian copula.
This has several interesting consequences, which are elucidated in \S 4.
Since our main purpose in this paper is to introduce
this technique into the field of large-scale structure, we defer more detailed
investigations to a later paper.


\section{What is a copula?} \label{sec:copintro}

The discussion in this section is taken primarily from
Nelson (1999), Malevergne \& Sornette (2003), and Genest \& Favre (2007).
Note that the terminology
in the statistics literature tends to differ slightly from that
used in cosmology; we will use the latter terminology here.

Consider the PDF of the distribution of densities
at $n$ points, ${\bf r_1}$, ${\bf r_2},...{\bf r_n}$.  We will denote
this $n$-point PDF as $p_n(\delta_1, \delta_2,...,\delta_n)$.  As noted
in the previous section, a great deal of work has been devoted to the
investigation of the 1-point distribution, $p(\delta)$.
The copula is a function that provides all of the remaining information necessary
to construct the $n$-point PDF, once this 1-point PDF is known.  Hence,
it couples together the individual 1-point PDFs to produce the full $n$-point
PDF; this is the origin of the term ``copula."  Since the
statistics of density fields in large-scale structure are translation-invariant,
all of our 1-point PDFs will be identically the same, but this need
not be the case for the general definition of the copula.

The copula is defined in terms of the $n$-point
cumulative distribution function (CDF) rather than
PDF.  Recall that the $n$-point CDF, $P_n(\delta_1, \delta_2,... \delta_n)$
is defined as:
\begin{eqnarray}
P_n(\delta_1, \delta_2,... \delta_n) = \int_{-\infty}^{\delta_1}
\int_{-\infty}^{\delta_2}
...\int_{-\infty}^{\delta_n} p(\widetilde \delta_1,
\widetilde \delta_2,...\widetilde \delta_n)\nonumber\\
\times d\widetilde\delta_1 d\widetilde\delta_2...
d\widetilde\delta_n,
\end{eqnarray}
and the definition of the 1-point CDF is just
\begin{equation}
P(\delta) = \int_{-\infty}^\delta p(\widetilde \delta) d\widetilde\delta.
\end{equation}
(We follow the standard convention of lower-case symbols for PDFs and upper-case
symbols for CDFs).
Then the copula function $C(u_1,u_2,...u_n)$ is the unique function that satisfies
the relation
\begin{equation}
\label{copdef}
P_n(\delta_1, \delta_2,...\delta_n) = C(P(\delta_1),P(\delta_2),...P(\delta_n)).
\end{equation}
Since we are describing a cosmological density field,
we can take
all of the 1-point CDFs on the right-hand side to be the same, but this is not
the most general definition of the copula.
Sklar's (1959) theorem states that a function
satisfying equation (\ref{copdef}) always exists, and that it is unique.
Hence, the $n$-point copula and the 1-point PDF
completely characterize the $n$-point PDF of the density field.

It might appear that we have gained nothing from this exercise, since we have simply
replaced an infinite hierarchy of correlation functions with an infinite hierarchy
of copula functions.  However, this is not the case.  The $n$-point copula function
contains significantly
more information than the corresponding $n$-point correlation function.
In the next section, for example, we characterize the 2-point copula for a simulated
evolved density field.  The information in the 2-point copula, along with the
1-point PDF, completely characterizes the 2-point density distribution function,
$p(\delta_1, \delta_2)$, which cannot be determined solely from a knowledge of the
2-point correlation function and the 1-point PDF.  A number of interesting conclusions
can be drawn from the 2-point copula alone.

Since CDFs vary between 0 and 1, the copula function maps an $n-$dimensional unit
cube onto the unit interval.  From the general properties of CDFs,
it follows that $C(u_1, u_2,...u_n) = 0$
when any single $u_i$ is 0, and $C(1, 1,...u_i,...1) = u_i$.

The
copula has an additional important property that we will exploit several times.  Consider
a density field $\delta_1, \delta_2,...\delta_n$, and a second density
field obtained by a local monotonic transformation on the first one:
$f_1(\delta_1), f_2(\delta_2),...f_n(\delta_n)$.  Then these two
density fields have the same copula.  Note that the functions
$f_1$, $f_2,...f_n$ do not have to be the same; all that is required is that
each function be a monotonic increasing function.  For instance, suppose we begin
with a Gaussian density field and exponentiate each $\delta$ to produce a log-normal
density field (Coles \& Jones 1991).  Then the initial Gaussian density field
and the corresponding log-normal density field have the same copula; the difference
between them is determined entirely by the 1-point PDF.

For simplicity, we will now confine our attention to 2-point copulas,
$C(u,v)$, with $0 \le u \le 1$, $0 \le v \le 1$, and $0 \le C(u,v) \le 1$.
There are several
special cases of interest.  First consider the
case of two uncorrelated densities, $\delta_1$ and $\delta_2$.  In this case,
$p(\delta_1, \delta_2) = p(\delta_1)p(\delta_2)$, so the copula is just
\begin{equation}
\label{uncor}
C(u,v) = uv.
\end{equation}
Since we will be dealing with Gaussian initial conditions, a second important copula
will be the Gaussian copula (see, e.g., Malevergne \& Sornett 2003) given by:
\begin{equation}
\label{gauss}
C_r(u,v) = \Phi_r[\Phi^{-1}(u),\Phi^{-1}(v)].
\end{equation}
Here $\Phi_r$ is the 2-point Gaussian CDF with unit variance and correlation $r$:
\begin{eqnarray}
\Phi_r(\delta_1,\delta_2) &=& \frac{1}{2\pi\sqrt{1-r^2}}\nonumber\\
\label{gaussr}
\times\int_{-\infty}^{\delta_1} \int_{-\infty}^{\delta_2}
&\exp&\left(  - \frac{1}{2(1-r^2)}(\delta_1^2+\delta_2^2 - 2r\delta_1\delta_2)\right),
\end{eqnarray}
while $\Phi^{-1}$ is the inverse of the 1-point Gaussian CDF with unit variance.

A Gaussian density field (such as that assumed for the initial conditions for large-scale
structure) has both a Gaussian copula and a Gaussian 1-point distribution.  However,
it is possible for a non-Gaussian density field to have a Gaussian copula (e.g., any
local monotonic transformation on a Gaussian field, such as the lognormal model discussed
above), and it is also possible for a field to have a Gaussian 1-point distribution
and a non-Gaussian copula.  In the latter case, the copula formalism provides a convenient
way to generate a variety of non-Gaussian fields with Gaussian 1-point PDFs (Nelson 1999).


\section{The Copula of the Nonlinear Density Field} \label{nonlinear}

Armed with the results of the previous section, we now examine an
evolved nonlinear density field.  Using the standard $\Lambda$CDM model, 
we analyze the mass distribution from a high resolution $N$-body 
simulation from the LasDamas project (McBride et al., in prep). The simulation 
was run with $1400^3$ particles in a box of side length $420 \hmpc$, and a 
flat cosmology specified by $\Omegam = 0.25$, $\Omega_\Lambda = 0.75$, 
$H_o = 70\;\kms\;{\rm Mpc}^{-1}$, $\sigma_8 = 0.8$, $n_s = 1.0$. 
We sample the density field at redshift zero using a spherical tophat of radius $1\hmpc$, corresponding
to a highly nonlinear density field.  Given the resolution 
of the simulation, the mean number of particles per
sphere is 160. The evolved 1-point PDF is shown in 
Fig. 1; it is highly non-Gaussian.
\begin{figure}[t]
	\epsfig{file=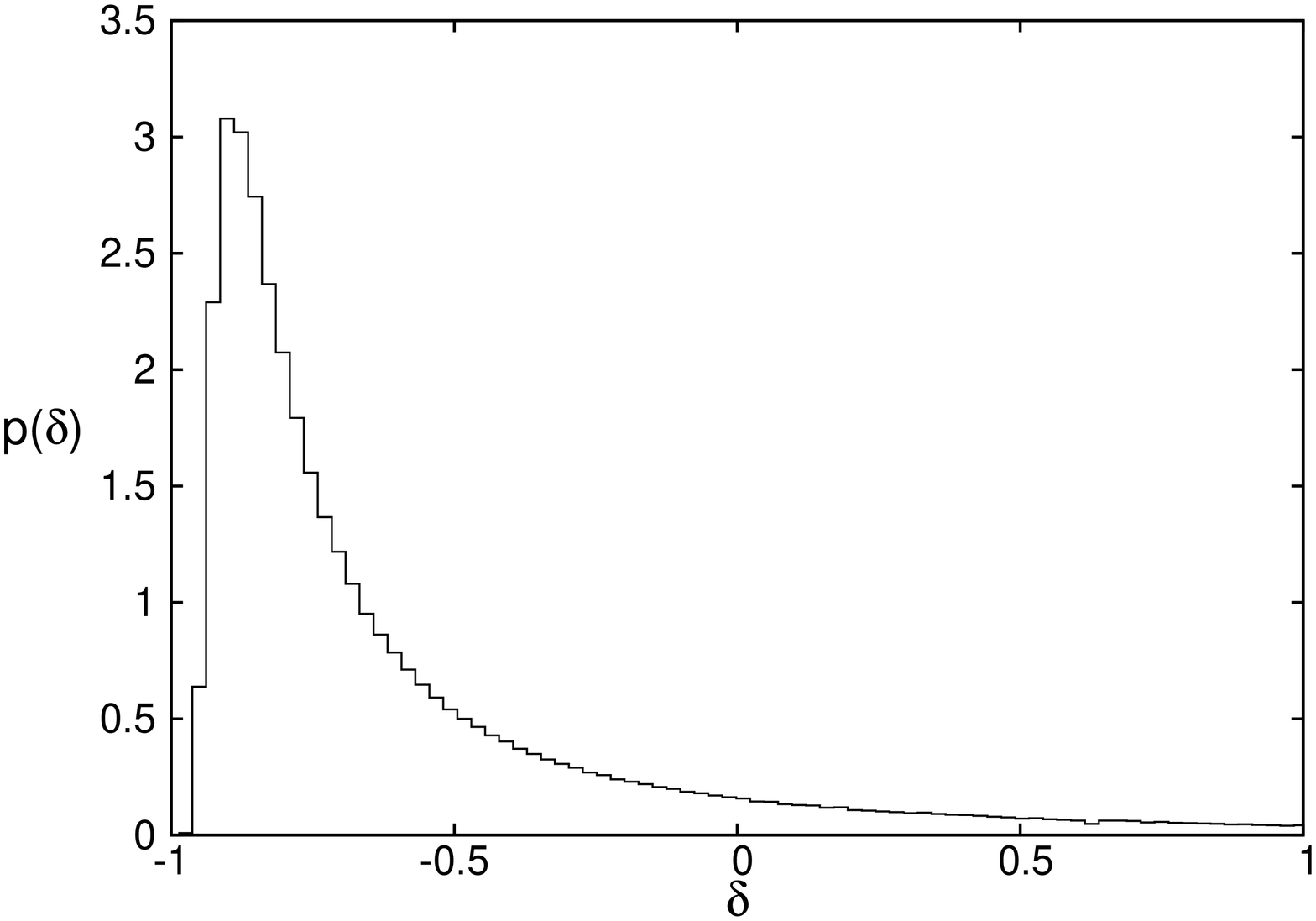,height=3.5in,angle=270}
	\caption
	{The 1-point PDF of our density field, sampled with a spherical tophat
	window function of radius $1\hmpc$.  
	}
\end{figure}
To determine the 2-point copula, we sample pairs of points separated by
$2 \hmpc$ and $6 \hmpc$, respectively.  Our goal
is to measure the copula for both $\xi < 1$ and $\xi > 1$, and we find that
the 2-point correlation of dark matter particles at these separations
is $\xi(2\hmpc) = 6.63$ and $\xi(6\hmpc) = 0.873$.  At much larger 
separations, where $\xi \ll 1$, the densities at the two points are 
essentially uncorrelated, and the copula simply takes the form in Equation (\ref{uncor}).

We sample 163,216 pairs of densities at each of the two separations.  We then use these
density pairs to derive the ``empirical copula", using the procedure outlined in
Genest \& Favre (2007).  We exploit the fact that the
copula is unchanged if we make a local monotonic transformation on the density field.
The particular monotonic transformation we make on each of our two columns of densities is
to replace each density by its rank within its own column, $R(\delta_i)$.
Thus, a given density pair, $\delta_1, \delta_2$, is mapped to $R_1(\delta_1),
R_2(\delta_2)$, where the ranking is determined separately for each column of
densities.  Then we divide by the number of pairs of points, $n = 163,216$, to
give $R_1(\delta_1)/n, R_2(\delta_2)/n$.  It is easy to see that the
distribution of ranks divided by the number of ranked points has a uniform
CDF.  Hence, for our new distribution, the right-hand side
of Equation (\ref{copdef}) has $P(R_1(\delta_1)/n) = R_1(\delta_1/n)$,
$P(R_2(\delta_2)/n) = R_2(\delta_2/n)$, and the equation becomes
\begin{equation}
P(R_1(\delta_1)/n,R_2(\delta_2)/n) = C(R_1(\delta_1)/n,R_2(\delta_2)/n).
\end{equation}
In other words, the 2-point distribution obtained by replacing each density
with its rank (divided by the number of points) {\it is} the 2-point copula.
The copula obtained in this way is called the empirical copula.

We have used our sampled pairs of points to derive the empirical copula for both
separations.  Since the 2-point copula is a mapping from $[0,1] \times [0,1]$
into $[0,1]$, we have chosen to display the copulas as contour plots in Figs.
2 and 3.  This empirical 2-point copula, displayed as a solid contour, is
the main result of this paper; along with the 1-point PDF for the
density, it provides a complete description of the 2-point density
distribution at the given separation.

However, we can go further and ask if the empirical copula corresponds
to any simple functional behavior.  Since the initial copula is Gaussian,
the obvious choice is the Gaussian copula given by Equation (\ref{gauss}).
This raises an obvious question:  what value of $r$ do we assume for our
theoretical Gaussian copula?  This value of $r$ will {\it not}, in general, correspond
to the normalized 2-point correlation function of the density field,
$\xi/\sigma^2$, since the latter also depends on the specific
1-point PDF.  Instead, we follow Genest \& Favre (2007) to
compute Spearman's $\rho$ for the data, and convert this into the value of $r$
for a corresponding Gaussian.

Spearman's $\rho$ is essentially the correlation function for the data ranks.  Let
$R_{1i}$ and $R_{2i}$ be the ranks of the $i$th data point in each of our two columns
of data.  Then Spearman's $\rho$ for our $n$ pairs of data points is defined as
\begin{equation}
\rho = \frac{\sum_{i=1}^n(R_{1i} - \bar R)(R_{2i} - \bar R)}{\sqrt{(\sum_{i=1}^n(R_{1i}-\bar R)^2)
(\sum_{i=1}^n(R_{2i}-\bar R)^2)}}.
\end{equation}
Here $\bar R$ is the mean value of the rank, which is, of course, $\bar R = (n+1)/2$.
The value of $\rho$ is related to an integral over the copula (Nelson 1999; Genest \& Favre 2007):
\begin{equation}
\rho = 12 \int_{u=0}^1 \int_{v=0}^1 C(u,v) du dv - 3.
\end{equation}
For a Gaussian copula, the relation between Spearman's $\rho$ and the value of $r$
that appears in Equation (\ref{gaussr}) is (Kruskal 1958; Genest \& Favre 2007)
\begin{equation}
r = 2 \sin(\pi \rho/6).
\end{equation}
The values of $\rho$ for our data are
$\rho(2\hmpc) = 0.474$ and $\rho(6\hmpc) = 0.139$, which correspond to
$r(2\hmpc) = 0.491$ and $r(6\hmpc) = 0.146$.  Using these values for $r$,
and equations (\ref{gauss})-(\ref{gaussr}), we have constructed the Gaussian copulas that
should provide the best fit to the empirical copulas, if the latter are indeed Gaussian.
These are displayed in Figs. 2-3.  The Gaussian copulas appear to match the empirical
copulas in both cases.

\begin{figure}[t]
	\epsfig{file=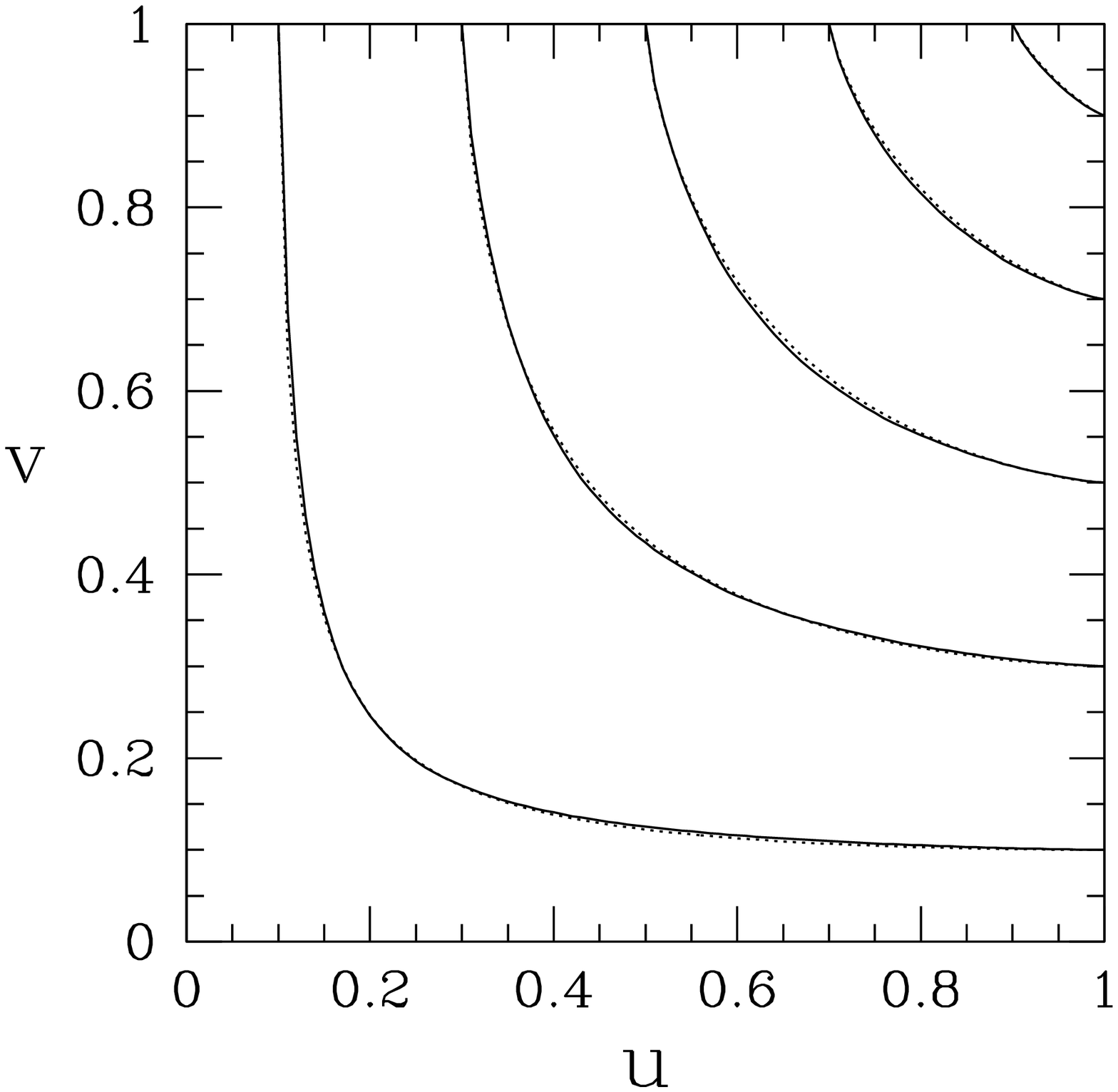,height=3.5in}
	\caption
	{The empirical 2-point copula $C(u,v)$ for a simulated dark matter density
	distribution at a separation of $2\hmpc$.  Solid curves
	are the contours corresponding to (from lower left to upper right)
	$C(u,v) = 0.1, 0.3, 0.5, 0.7, 0.9$.
	Dashed curves give the Gaussian copula with the value of $r$
	corresponding to Spearman's $\rho$ calculated for the data.
	}
\end{figure}

\begin{figure}[t]
	\epsfig{file=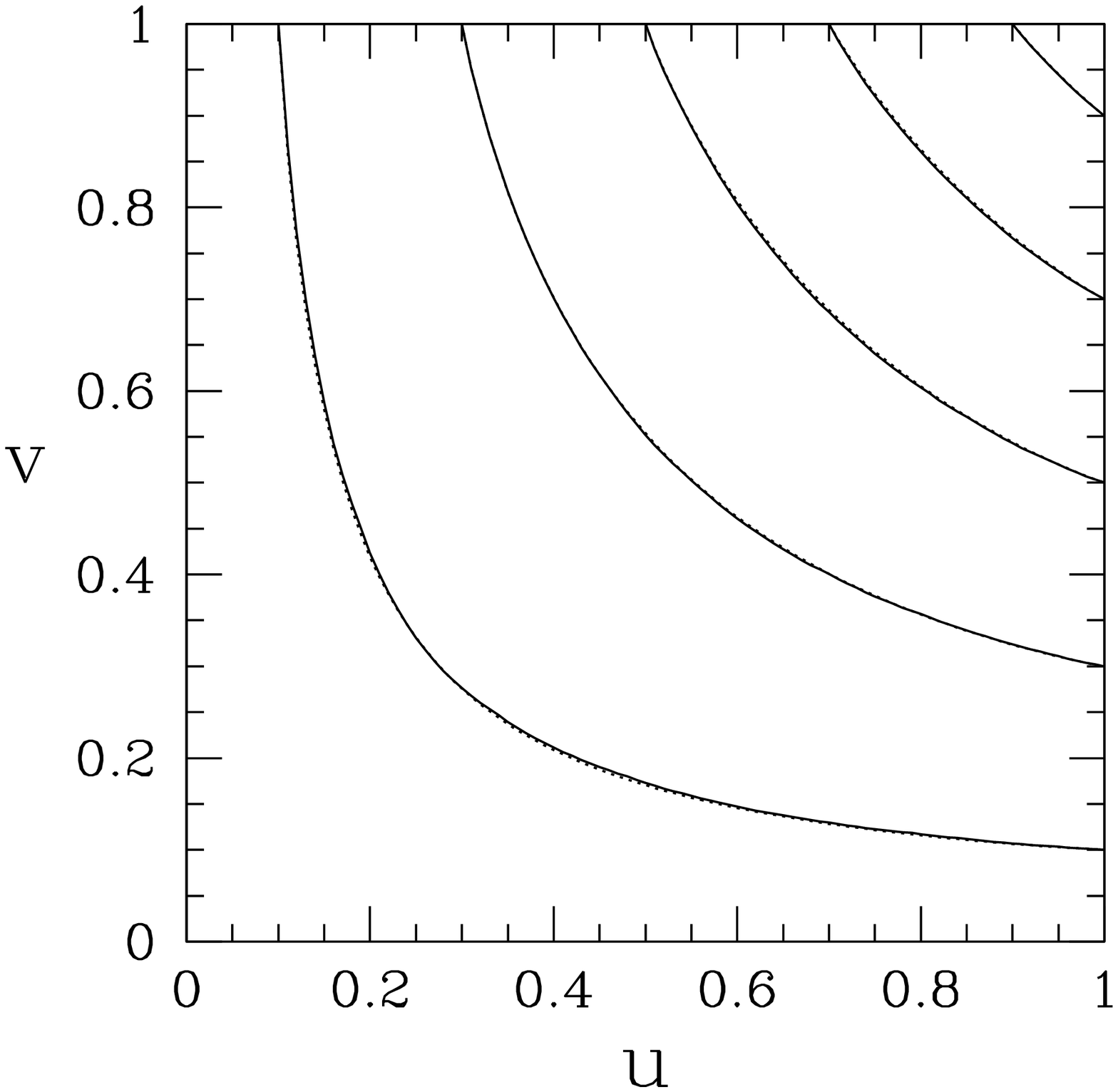,height=3.5in}
	\caption
	{As Fig. 2, for a separation of $6 \hmpc$.  
	}
\end{figure}

\section{Discussion} \label{sec:discussion}

Our results indicate that the two-point copula for the present-day dark matter density field
is well-approximated by a Gaussian copula.  This result, along with a knowledge of the 1-point
PDF, is sufficient to completely characterize $p(\delta_1,\delta_2)$.
The most obvious open question is then whether all of the higher-order copulas are also Gaussian;
we will defer investigation of this Gaussian copula hypothesis (GCH) to a future paper.
If the GCH were true,
it would imply that the nonlinear density field could be derived
by a local transformation of an underlying Gaussian field, an idea which has
been explored in the past (see, e.g., Coles \& Jones 1991).  Note, however,
that this does {\it not} imply that the evolved density field is a local transformation
of the {\it original} Gaussian dark matter density field; the Gaussian field that is locally
mapped to produce the final density field could be some other Gaussian density field.
But the GCH would imply that all of the non-Gaussian information in the nonlinear density field
could be derived in terms of the 1-point PDF.  For example, all of the higher-order correlation
functions would depend only on this PDF.

These arguments are related to the Gaussianization process of Weinberg (1992).  Weinberg explored
the possibility that gravitational evolution preserves the rank order of the density field,
so that mapping the nonlinear density field monotonically onto a Gaussian field would reproduce
the initial density field.  It is clear that this process changes only the 1-point PDF and
leaves the copula unchanged.  The results on reconstruction were somewhat mixed; while there
is a reasonable correlation between the initial density field and the reconstructed density field,
the correspondence is certainly not exact (Narayanan \& Croft 1999).  However, this result
does not contradict the GCH; as noted above, there is no reason to assume that the Gaussian
field that is locally-transformed into the final density field is identical to the initial
Gaussian density field. In fact the two Gaussian fields could even
have different values for $r$.  (See also the discussions of Pando, Feng, \&
Fang 2001 and Neyrinck, Szapudi, \& Szalay 2009 on these issues).

A more direct constraint on the GCH comes from measures of topology (Doroshkevich 1970; Hamilton, Gott, \&
Weinberg 1986; Gott, Weinberg, \& Melott 1987; Weinberg, Gott, \& Melott 1987;
Melott, Weinberg, \& Gott 1988), or more generally,
Minkowski functionals (Mecke, Buchert, \& Wagner 1994; Kerscher, et al. 1997).
When the independent
variable in these calculations is taken
to be the volume filling factor, rather than the density
threshold, then such statistics
effectively divide out the effect of the 1-point PDF; therefore,
they can depend only on the behavior of the copula (see, e.g., Shandarin 2002 for
a detailed discussion of this point).  For the case of topology, the GCH then implies that
the genus curve of the nonlinear evolved density field will have the shape characteristic
of a Gaussian density field (unlike the case of Gaussianization, this result does not depend
on the Gaussian copula matching the initial Gaussian density field).  This was claimed
to be the case in the first simulations of topology (Weinberg, Gott, \& Melott 1987; Melott, Weinberg, \&
Gott 1988).  More recent simulations (Park, Kim, \& Gott 2005; Kim et al.
2009) indicate that the genus curve retains its Gaussian shape for moderate
smoothing lengths, but it clearly departs from Gaussianity (in terms
of the ``shift parameter", which is the relevant quantity here) on
the highly nonlinear length scale we have examined ($1\hmpc$).
These results argue against the GCH on nonlinear scales.
Clearly, the higher-order
copula functions are worthy of further study.

Of course, we actually observe the distribution of galaxies, and not dark
matter.  The discussion in the previous sections shows that for biasing
schemes that are local and monotonic (such as those explored by
Coles 1993; Fry \& Gaztanaga 1993; Scherrer \& Weinberg 1998; Coles,
Melott, \& Munshi 1999; Narayanan, Berlind, \& Weinberg 2000) the copula of the galaxy distribution will be
identical to the copula of the underlying dark matter density field.
This will not necessarily be the case for nonlocal bias, or stochastic bias 
(Dekel and Lahav 1999).  The best current models include some degree of
stochastic bias; what remains to be seen is the size of the effect on the
copula.

This short introductory paper leaves open a number of questions, several
of which we are currently
investigating.  The most important is whether the higher-order copulas of the density field
are also Gaussian.  While it is obviously impossible to examine this question to all orders,
an investigation of the 3-point copula is straightforward and should provide a useful check.
Other directions for future investigation are the effects of nonlocal or
stochastic bias, redshift distortions, and the copula of the observed galaxy
distribution.

We believe that the copula has the potential to serve
as an important new tool in the analysis of large-scale structure.  It appears
to be less sensitive to bias (e.g., completely unaffected by local monotonic
bias) than
other statistics.  If the GCH applies, then the full density field can be completely
characterized by a single function (the 1-point PDF) and a series of numerical
parameters (the correlations $r$ for the copula as a function of length
scale).  For example, in this case the hierarchical clustering coefficients can be derived
as functions of the 1-point PDF.
Even if the GCH does not apply, the copula allows us to measure
the underlying ``coupling" between the density field at different points
in an entirely new way, moving beyond the limited information in the low-order
correlation functions.  The copula can also be used to
analyze the evolution of the density field, via a computation of the
two-point copula for
the density measured at the same points in the initial and final density fields. 

We note in passing that it is precisely the Gaussian copula which
has been blamed for the recent mortage-backed securities meltdown.  We presume
than any error in this paper will have less dire consequences.


\acknowledgments 

R.J.S. was supported in part by the Department of Energy (DE-FG05-85ER40226)
We thank D.H. Weinberg for helpful discussions.



\def\baselinestretch{1}

\end{document}